\documentclass[a4paper,12pt]{article}
\usepackage{amsmath}
\usepackage{amssymb}
\usepackage{xfrac}
\usepackage{verbatim}
\usepackage{graphicx}
\usepackage[CaptionAfterwards]{fltpage}
\usepackage[font=small]{caption}
\usepackage{subcaption}
\usepackage{pdfpages}
\usepackage{textcomp}
\usepackage{bibentry}
\nobibliography*

\newcommand{\term}[1]{\textit{#1}}
\newcommand{\ct}[1]{$\mathrm{#1}$}
\newcommand{\abs}[1]{\left|{#1}\right|}

\newcommand{\ket}[1]{\left|{#1}\right\rangle}
\newcommand{\bra}[1]{\left\langle{#1}\right|}
\newcommand{\braket}[2]{\left\langle{#1}\right| \left.{#2} \right\rangle}

\newcommand{\colvec}[2]{\begin{pmatrix}{#1}\\{#2}\end{pmatrix}}

\renewcommand{\phi}{\varphi}

\title{An Electronic Quantum Eraser\\Supplemental Material}
\author{E. Weisz, H. K. Choi , I. Sivan, M. Heiblum, Y. Gefen,\\ D. Mahalu, and V. Umansky }
\date{}

\begin{document}


# An Electronic Quantum Eraser

E. Weisz[*], H. K. Choi[*], I. Sivan, M. Heiblum[†], Y. Gefen, D. Mahalu, and V. Umansky

Braun Center for Submicron Research, Department of Condensed Matter Physics, Weizmann Institute of Science, Rehovot 76100, Israel.
[*] These authors contributed equally to this work.
[†] E-mail: moty.heiblum@weizmann.ac.il

# Introduction

Complementarity, the incomplete nature of a quantum measurement - a core concept in quantum mechanics - stems from the choice of the measurement apparatus[1]. The notion of complementarity is closely related to Heisenberg's *uncertainty principle*, but the exact relation between the two remains a source of debate[2-8]. For example, knowledge of a particle's position in a *double slit* interference experiment will quench its wave-like nature and, vice versa, observing the wave property via interference implies lack of knowledge of the particle's path. A canonical system for exploring complementarity is the *quantum eraser* (QE), predominantly studied thus far in photonic systems[9-14]. A QE is an interference experiment consisting of two stages. First, one of the interfering paths is coupled to a 'which path' (WP) detector - demonstrating loss of interference due to acquisition of WP information. Second, the WP information is being 'erased' by projecting the detector's wavefunction on a particular basis; this renders the WP information inaccessible, thus allowing reconstruction of the interference pattern.

In this work, we present a first implementation of a QE in an electronic system. Our system consists of two identical electronic Mach-Zehnder interferometers (MZIs)[15] entangled via Coulomb interactions. Such novel setup has already attracted a considerable theoretical attention[16-18]. With one MZI serving as a path detector and the other as the system interferometer, the visibility of the Aharonov-Bohm oscillation in the System can be controlled by the Detector. We demonstrate how a continuous change of the measurement basis, followed by *post selection* (via cross correlation of current fluctuations), allows a smooth transition between keeping and erasing the WP information.

# Theory
## Mach-Zehnder interferometer

An electronic MZI is formed by manipulating quasi one dimensional, chiral edge channels, which are formed in the integer quantum Hall effect regime[15]. Such a realization allows directing the path of electrons at will - leading to high visibility interference pattern. Potential barriers, formed by quantum point contacts (QPCs), take the role of optical beam splitters, transmitting and reflecting impinging electrons with amplitudes $t_i$ and $r_i$, respectively, with $|t_i|^2 + |r_i|^2 = 1$.

Two such coupled MZIs are shown in Fig. 1, where the coupling is mediated by the lower path of the System and the upper path of the Detector, referred to as *interacting paths* (shaded area in Fig. 1a).

Starting with the System, an electron injected from Source S1 arrives at SQPC1 and is put into a superposition of being reflected into the interacting path and transmitted into the non-interacting path, namely:

$$(1) \quad |S\rangle = r_1 |\uparrow\rangle_S + t_1 |\downarrow\rangle_S ,$$

with $|\uparrow\rangle_S$ and $|\downarrow\rangle_S$ standing for the two paths. The paths recombine and interfere at SQPC2, with the electron's probability of reaching the Drain D2 being:

$$(2) \quad P(\text{D2}) = |r_1 r_2 e^{i\phi_S} - t_1 t_2|^2 = T_0 - T_1 \cos(\phi_S),$$

where $T_0 \equiv |t_1 t_2|^2 + |r_1 r_2|^2, T_1 \equiv 2 t_1 t_2 r_1 r_2$, $\phi_S = 2\pi AB/\Phi_0$ is the Aharonov-Bohm (AB) phase[19]. Here, $\Phi_0 = h/e$ is the magnetic flux quantum, $A$ the area enclosed by the two paths, and $B$ the magnetic field. The visibility of the interfering pattern at D2 is defined as

$$(3) \quad v_{\text{D2}} \equiv \frac{\max[P(\text{D2})] - \min[P(\text{D2})]}{\max[P(\text{D2})] + \min[P(\text{D2})]} = \frac{T_1}{T_0}.$$

Throughout our experiments, all the QPCs were tuned to have equal transmission and reflection amplitudes, $|r_i|^2 = |t_i|^2 = \frac{1}{2}; i = 1\ldots 4$.

## Entangling two Mach-Zehnder interferometers

Coulombic coupling between the System and the Detector, with the Detector acquiring WP information of the System, and *vice versa*, should in effect lead to their entanglement[20-22]. The Coulomb interaction between two electrons passing simultaneously in the interacting paths causes a slight repulsion among them, affecting their trajectories and reducing the AB area in each MZI by $\delta A_\gamma$, consequently adding a phase shift $\gamma = \delta A_\gamma B/\Phi_0$ (Ref. 23). Hence, the states of the Detector are: $|0\rangle_D = \frac{1}{\sqrt{2}}(e^{i\phi_D}|\uparrow\rangle_D + |\downarrow\rangle_D)$, if the system's electron passes through the non-interacting path, and $|\gamma\rangle_D = \frac{1}{\sqrt{2}}(e^{i(\phi_D+\gamma)}|\uparrow\rangle_D + |\downarrow\rangle_D)$, if it passes through the interacting path. Consequently, the 'System – Detector' complex is now in an entangled state:

$$(4) \quad |\Psi\rangle = \tfrac{1}{\sqrt{2}}(|\downarrow\rangle_S |0\rangle_D + e^{i\phi_S}|\uparrow\rangle_S |\gamma\rangle_D).$$

The meaning of the entanglement here is that the Detector's state provides WP information about the System. Consequently the System is dephased, as evident by its transmission probability:

(5)
$$P(\text{D2}) = \tfrac{1}{2} + \tfrac{1}{2}\text{Re}(e^{i\phi_S}\langle 0|\gamma\rangle_D) = \tfrac{1}{2} + \tfrac{1}{4}[\cos(\phi_S) + \cos(\phi_S + \gamma)] = \tfrac{1}{2} + \tfrac{1}{2}\cos\left(\tfrac{\gamma}{2}\right)\cos\left(\phi_S + \tfrac{\gamma}{2}\right).$$

This can be understood as the average of two interference patterns of the System: one with the unperturbed AB phase, and one with the added interaction phase $\gamma$ (Fig. 2, inset). Indeed, tracing-out over the Detector states collapses the System into a statistical average of the electron traversing the interacting or the non-interacting paths. An equivalent view[24] stems from the realization that the Detector's interacting path carries shot noise (due to the current partitioning by DQPC1), which evidently leads to (at least partial) dephasing of the System.

3

The visibility of the System, an indicator of the its coherence, is determined by the overlap of the Detector states, $\langle 0|\gamma\rangle_D$:

$$(6) \quad v_{D2} = |\langle 0|\gamma\rangle_D| = \cos\left(\frac{\gamma}{2}\right).$$

Equivalently, the Detector states' *distinguishability D,* its efficacy in acquiring the WP information, is complimentary to the System's interference visibility, namely, $D^2 + v_{D2}^2 = 1$ (Ref. 25-27), hence $D = \sin\left(\frac{\gamma}{2}\right)$. For example, if the Detector's states are orthogonal for System electrons choosing the interacting *vs.* the non-interacting path (*i.e.* $\gamma=\pi$ and *D*=1), then *a priori* full dephasing is expected in the system ($v_{D2}$=0). In terms of complementarity, having the particle's WP information, encoded in the detector's phase, comes at the expense of observing the particle's wave-like nature.

## Joint detection

The fact that the WP information can be measured, even if 'in principle', is sufficient to (at least partly) dephase the interferometer, demonstrating its particle-nature. Can we recover the wave-like nature? Indeed, by interfering the two paths of the Detector at DQPC2, the measured current at drain D4 can be tuned to move between two extremes. At one extreme, it carries maximal WP information about the System – hence, the interference cannot be recovered. At the other extreme, it does not carry WP information anymore – hence, recovery of interference is possible (by post-selection of events).

One can define the *knowledge K,* the WP information measured in our Detector, at Drain D4:

$$(7) \quad K(\phi_D) \equiv |P(\text{D4}|\uparrow_S) - P(\text{D4}|\downarrow_S)| = \left|\tfrac{1}{2}[\cos(\phi_D + \gamma) - \cos(\phi_D)]\right| = \left|D\sin(\phi_D + \tfrac{\gamma}{2})\right|,$$

namely, the difference in the Detector's output currents, for a certain AB phase $\phi_D$, when the System's electron chooses either of the possible paths. While the distinguishability *D* indicates the WP information encoded in the Detector state, the knowledge *K* indicates how much of it is actually accessible. Naturally, the knowledge is bound by the distinguishability, $K \leq D$. The knowledge, *K*, can be continuously tuned by altering the detector's AB phase $\phi_D$, between gaining full WP information (*K*=*D*) and fully erasing it *(K*=0).

Erasing the WP information allows recovery of the electron's wave-like nature, manifested as AB interference. This can be accomplished by *post selecting* readings of the System's output according to the Detector's reading. Experimentally, we measure the correlation between the current fluctuations in Drains D2 (System) and D4 (Detector)[22, 28], proportional to the reduced joint-probability for simultaneously detecting electrons at the two drains (see supplemental material and Ref. 18),

$$(8)\ P(\delta D2 \times \delta D4) = \frac{1}{4}\cos\left(\phi_S + \frac{\gamma}{2}\right)\cos\left(\phi_D + \frac{\gamma}{2}\right)\sin^2\left(\frac{\gamma}{2}\right).$$

Note that the joint probability $P(\delta D2 \times \delta D4)$ is symmetric with regard to System and Detector, stressing the arbitrariness of their labeling. The expected visibility of the AB oscillation in the reduced joint probability is $v_{D2D4} = \sin\left(\frac{\gamma}{2}\right)\cos\left(\phi_D + \frac{\gamma}{2}\right)$, which also obeys a complementarity relation with the WP knowledge: $v_{D2D4}^2 + K^2 = D^2$ (Ref. 16).

# Results
## Dephasing

We realized the System-Detector complex in two dimensional electron gas (2DEG), embedded in GaAs-AlGaAs heterostructure. A quantizing magnetic field ($B$=4.4 T) put the 2DEG in filling factor 2 of the quantum Hall effect, with electron temperature of 12 mK. Only the outer chiral edge channels (the lowest spin split Landau level) participated in the interference and the interaction, while the inner edge channels were fully reflected at QPC0, which precede QPC1 of both MZIs, and thus played no active role. The outer edge channel, emanating from Source contact S1 (Detector contact S3), was split at SQPC1 (DQPC1) to two paths that subsequently interfered at SQPC2 (DQPC2). The AB phase is controlled by changing the area enclosed between the interfering paths via the System's modulation gate (SMG). Filling factor 2 was chosen since the devices are more stable and the visibility of the AB oscillation was reasonably high; however, the two interacting paths were partly screened by the adjoining (unbiased) inner channels – weakening thus the mutual interaction. The output current, $\langle I_D \rangle$ (in D2 and D4), was proportional to the phase-dependent transmission probability $P(D)$. For half transmission of QPC1 and QPC2, $|t_1|^2 = |t_2|^2 = \frac{1}{2}$, the visibility reached $v \approx 0.66$ in each MZI (with dephasing from uncontrolled charge fluctuations or electron interactions). Interaction between System and Detector via the interacting paths, counter-propagating along 6 μm and separated by less than 2 μm, was controlled by surface gates, avoiding tunneling between the interacting paths.

Under these conditions, the System's differential visibility $v_{D2}$ was measured as a function of the Detector's current $I_{S3}$ (Fig. 2). The visibility $v_{D2}$ decreased by some 15% with increasing the input current to 1 nA. This can be understood in two ways: (*i*) With the increased current, more electrons are packed at any time into the interacting path of the Detector, increasing $\gamma$ and consequently the acquired WP information; (*ii*) In terms of interaction, potential fluctuations in the Detector interacting path increase with current, thus leading to stronger dephasing of the System. For a linear dependence of $\gamma$ on $I_{S3}$ the data fits well Eq. 6, with $\gamma \approx \frac{\pi}{6}$ at $I_{S3}$=1 nA. Note that the transmission of DQPC2 was varied and was found to have no effect on the dephasing of the System – as expected.

## Autocorrelation (AC)

After quantifying the strength of interaction, equal source currents $I_S$=0.5 nA were fed into both MZIs. The interacting paths were set to obtain maximal interaction, void of inter channel tunneling. Three 'zero-frequency' correlation signals were measured simultaneously: (*i*, *ii*) Auto-correlation (AC) of each drain current $\langle \delta I_D^2 \rangle$; (*iii*) Cross-correlation (CC) of the two drain currents $\langle \delta I_{D2} \delta I_{D4} \rangle$. All three were measured as function of SMG, affecting the System's phase $\phi_S$, and the decaying magnetic field which affected both $\phi_S$ and $\phi_D$.

The auto-correlation of each interferometer is proportional to its shot noise; In the case of the System:

$$(9) \quad \langle \delta I_{D2}^2 \rangle \propto S_{MZI} = 2eI_{S1}P(D2)[1 - P(D2)] = \tfrac{1}{4}eI_{S1}[1 - \cos(2\phi_S)],$$

where $P(D2)$ is the average transmission of the System. The shot noise oscillated at half the periodicity of the conductance (see Figs. 3b and 3c). The color plot in Fig. 3a was obtained by letting the magnetic field decay slowly while the voltage of SMG was scanned faster. Interestingly, the magnitude of the shot noise in each MZI was considerably larger than predicted by Eq. 9, a phenomenon that was previously observed[22]. Though this phenomenon remains unexplained, it could result from unobservable high frequency charge fluctuations (and thus in the AB phase), which lower the visibility right from the start – being down converted to the measured frequency by the partitioning of QPC2.

## Cross-correlation (CC)

The main result lies in the AB dependence of the zero-frequency CC between current fluctuations in D2 and D4, $\langle \delta I_{D2} \delta I_{D4} \rangle$. The CC signal, merely 4·10$^{-30}$A$^2$/Hz, was extracted by a digital band-pass filter and compensated during analysis for the unavoidable magnetic field dependence of the System's interference pattern (Fig. 4a).

It may be easier to understand the CC color plot with two cuts at certain magnetic fields (Fig. 4b): One cut at $\Delta B$=103μT (around $B$=4.4T), with relatively strong oscillations in the CC as function of $V_{SMG}$, and another at $\Delta B$=140μT, with reduced CC oscillations. The AB oscillations in the CC retrace qualitatively the *lost interference*; the dependence of the CC visibility on $\Delta B$ is plotted in Fig. 4c.

In a more quantitative fashion, the visibility of the CC oscillation (a wave-like property) is compared with the WP knowledge $K$ (a particle-like property), which is obtained by an approximation of Eq. 7 for $\gamma \ll \pi$, $K(\phi_D) \propto \left| \gamma \dfrac{dI_{D4}}{d\phi_D} \right|$. Although being readily available from the AB conductance oscillation in the Detector, we obtained the above derivative from the auto-correlation signal $\langle \delta I_{D4}^2 \rangle$, which was measured simultaneously with the CC signal. Plotting the CC visibility (Fig. 4c) and the knowledge $K$ (Fig. 4d), their anti-correlation dependence on $\Delta B$ is evident, a consequence of the principle of complementarity. The small shift between the two

dependencies is attributed to the finite $\gamma$, which we estimate based on the shift to be $\gamma \approx \frac{\pi}{12}$ (for $I_{S3}$=0.5 nA); this is consistent with the previously found $\gamma \approx \frac{\pi}{6}$ (for $I_{S3}$=1 nA). The oscillations in the CC visibility are larger than the ideal $\sin^2\left(\frac{\gamma}{2}\right) = 0.017$, similar to the excess noise in the AC. Reiterating again, having access to WP information suppresses the possibility to extract phase information, and vice versa.

## Epilogue

While this work proves an inevitable consequence of quantum mechanics, complementarity, it was performed with a novel *electronic quantum eraser* (EQE), a first of its kind. A major difference between electrons and photons or atoms, which are customarily used in this type of experiments, is their exchange statistics and strong Coulomb interaction. The first allows noise-free current (due to Pauli principle) and a highly controlled autocorrelation; the latter makes the electrons vulnerable to dephasing, but at the same time allows control of the entanglement strength and phase manipulation. This configuration is different from its optical-equivalent, featuring full System - Detector symmetry; hence, the labels of 'System' and 'Detector' are arbitrary. Future experiments could expand on the EQE, for example, by realizing a time delayed choice EQE, a spin-entangled EQE[29], or a three-electron entangled state (GHZ state)[30]. Moreover, proposals were made for utilizing the EQE to measure weak-values[31] or the Bell inequality[32] for the entangled electrons.

## Methods

### Device

We implemented a System-Detector complex, consisting of two coupled MZIs, in a GaAs-AlGaAs heterostructure harboring a two-dimensional electron gas (2DEG) at a depth of 63 nm below the surface. The 2DEG mobility was 2.5×10$^6$ cm$^2$/V·s and its electron density 2.9×10$^{11}$ cm$^{-2}$. The device boundaries were defined by applying negative voltage bias to Ti/Au surface gates; Ge/Au/Ni ohmic contacts provided access to the 2DEG. The interferometers were fabricated utilizing electron beam lithography.

### Low temperature

All the measurements took place in a dilution refrigerator, at a temperature of 12 mK. The sample was placed in a strong perpendicular magnetic field (4.4 T) generated by a super-current, decaying slowly over time. At these conditions the electron states in the 2DEG formed Landau levels, putting the device in the second filling factor of the integer quantum Hall effect regime. Two chiral edge channels formed at the crossing of the occupied Landau levels and the Fermi energy, each channel having a precisely known conductance of $e^2/h$.

### Cold amplifier

The device conductance was measured by applying an AC signal 1.5 µV at 730 kHz at the source contacts and probing the voltage at the drain contacts. The measured signal was amplified by a

two stage amplification chain, consisting of an in-situ *cold amplifier* (at 4.2K) and a room temperature amplifier, with a total voltage gain *g*=2000. The *zero-frequency* autocorrelations and cross-correlation were measured simultaneously by injecting DC current at the source contacts and measuring the resulting voltage fluctuations at 730 kHz over a bandwidth of 50 kHz. The signals were amplified, multiplied and passed through a low-pass filter in a custom-made analog electronic setup.

## Acknowledgements

# References

[1] Bohr, N. The quantum postulate and the recent development of atomic theory. *Nature* **121**, 580–591 (1928).

[2] Feynman, R., Leighton, R. and Sands, M. Quantum Mechanics, The Feynman Lectures on Physics, Vol. 3 *Addison-Wesley, Massachusetts* (1965).

[3] Scully, M.O., Englert, B.G. and Walther, H. Quantum optical tests of complementarity. *Nature* **351**, 111–116 (1991).

[4] Englert, B.G., Scully, M.O. and Walther, H. Complementarity and uncertainty. *Nature* **375**, 367–368 (1995).

[5] Storey, E.P. et al. Complementarity and uncertainty. *Nature* **375**, 368–368 (1995).

[6] Chang, D.I. et al. Quantum mechanical complementarity probed in a closed-loop Aharonov–Bohm interferometer. *Nature physics* **4**, 205–209 (2008).

[7] Dürr, S., Nonn, T. and Rempe, G. Origin of quantum-mechanical complementarity probed by a 'which-way' experiment in an atom interferometer. *Nature* **395**, 33–37 (1998).

[8] Bertet, P. et al. A complementarity experiment with an interferometer at the quantum–classical boundary. *Nature* **411**, 166–170 (2001).

[9] Scully, M.O. and Drühl, K. Quantum eraser: A proposed photon correlation experiment concerning observation and "delayed choice" in quantum mechanics. *Phys. Rev. A* **25**, 2208–2213 (1982).

[10] Kwiat, P.G., Steinberg, A.M. and Chiao, R.Y. Observation of a ''quantum eraser'': A revival of coherence in a two-photon interference experiment. *Phys. Rev. A* **45**, 7729 (1992).

[11] Herzog, T.J. et al. Complementarity and the quantum eraser. *Phys. Rev. Lett.* **75**, 3034–3037 (1995).

[12] Kwiat, P.G., Steinberg, A.M. and Chiao, R.Y. Three proposed ''quantum erasers''. *Phys. Rev. A* **49**, 61 (1994).

[13] Kim, Y.H. et al. Delayed "choice" quantum eraser. *Phys. Rev. Lett.* **84**, 1–5 (2000).

[14] Ma, X.S. et al. Quantum erasure with causally disconnected choice. *Proceedings of the National Academy of Sciences* **110**, 1221–1226 (2013).

[15] Ji, Y. et al. An Electronic Mach-Zehnder Interferometer. *Nature* **422**, 415-418 (2003).

[16] Kang, K. Electronic mach-zehnder quantum eraser. *Phys. Rev. B* **75**, 125326 (2007).

[17] Dressel, J., Choi, Y. and Jordan, A.N. Measuring which-path information with coupled electronic Mach-Zehnder interferometers. *Phys. Rev. B* **85**, 045320 (2012).

[18] Buscemi, F., Bordone, P. and Bertoni, A. Electron interference and entanglement in coupled 1D systems with noise. *The European Physical Journal D* **66**, 1–11 (2012).

[19] Aharonov, Y. and Bohm, D. Significance of Electromagnetic Potentials in the Quantum Theory. *Phys. Rev.* **115**, 485 (1959).

[20] Buks, E. et al. Dephasing in electron interference by a 'which-path' detector. *Nature* **391**, 871-874 (1998).

[21] Sprinzak, D. et al. Controlled Dephasing of Electrons via a Phase Sensitive Detector. *Phys. Rev. Lett.* **84**, 5820–5823 (2000).

[22]     Neder, I. et al. Entanglement, Dephasing, and Phase Recovery via Cross-Correlation Measurements of Electrons. *Phys. Rev. Lett.* **98**, 036803 (2007).

[23]     Rosenow, B. and Gefen, Y. Dephasing by a Zero-Temperature Detector and the Friedel Sum Rule. *Phys. Rev. Lett.* **108**, 256805 (2012).

[24]     Stern, A., Aharonov, Y. and Imry, Y. Phase uncertainty and loss of interference: A general picture. *Phys. Rev. A* **41**, 3436–3448 (1990).

[25]     Wootters, W.K. and Zurek, W.H. Complementarity in the double-slit experiment: Quantum nonseparability and a quantitative statement of Bohr's principle. *Phys. Rev. D* **19**, 473–484 (1979).

[26]     Jaeger, G., Shimony, A. and Vaidman, L. Two interferometric complementarities. *Phys. Rev. A* **51**, 54–67 (1995).

[27]     Englert, B.G. Fringe Visibility and Which-Way Information: An Inequality. *Phys. Rev. Lett.* **77**, 2154–2157 (1996).

[28]     Neder, I. et al. Interference between two indistinguishable electrons from independent sources. *Nature* **448**, 333-337 (2007).

[29]     Chen, W. et al. Quantitatively probing two-electron entanglement with a spintronic quantum eraser. *Phys. Rev. B* **87**, 155308 (2013).

[30]     Vyshnevyy, A.A. et al. Setup of three Mach-Zehnder interferometers for production and observation of Greenberger-Horne-Zeilinger entanglement of electrons. *Phys. Rev. B* **87**, 165417 (2013).

[31]     Shpitalnik, V., Gefen, Y. and Romito, A. Tomography of Many-Body Weak Values: Mach-Zehnder Interferometry. *Phys. Rev. Lett.* **101**, 226802 (2008).

[32]     Vyshnevyy, A.A. et al. Two-particle entanglement in capacitively coupled Mach-Zehnder interferometers. *Phys. Rev. B* **87**, 165302 (2013).

a

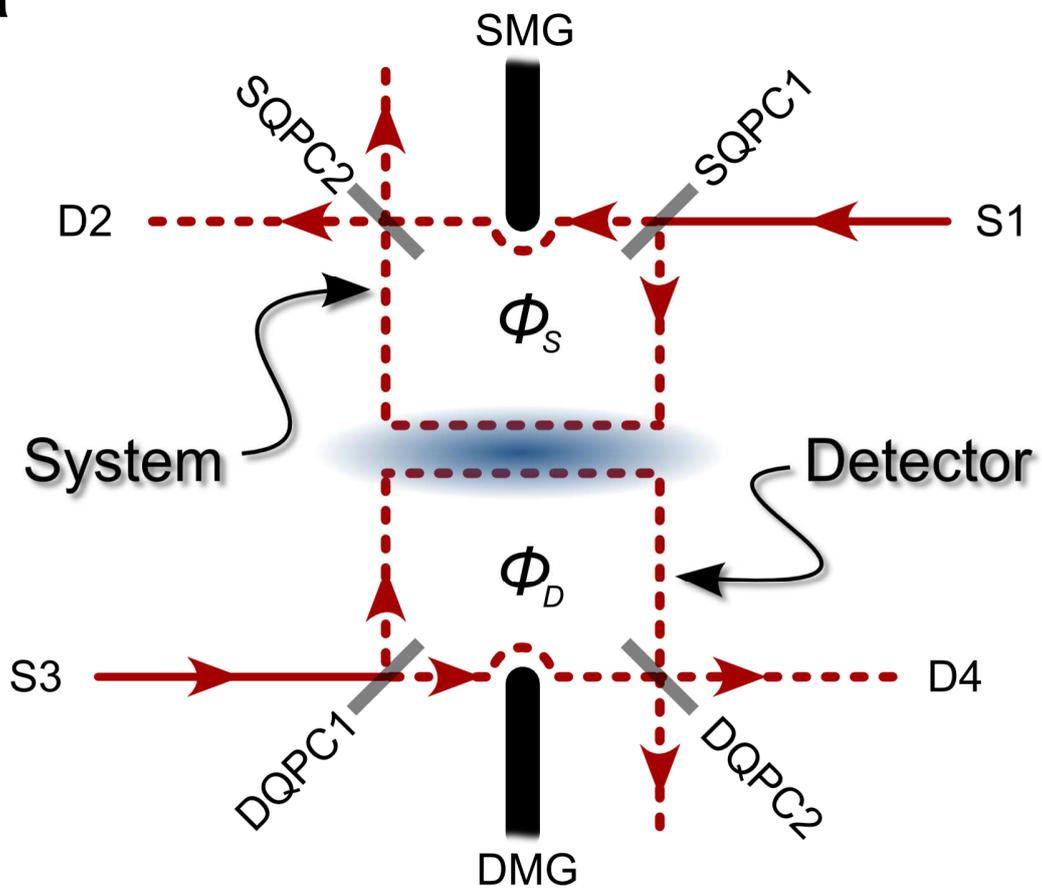

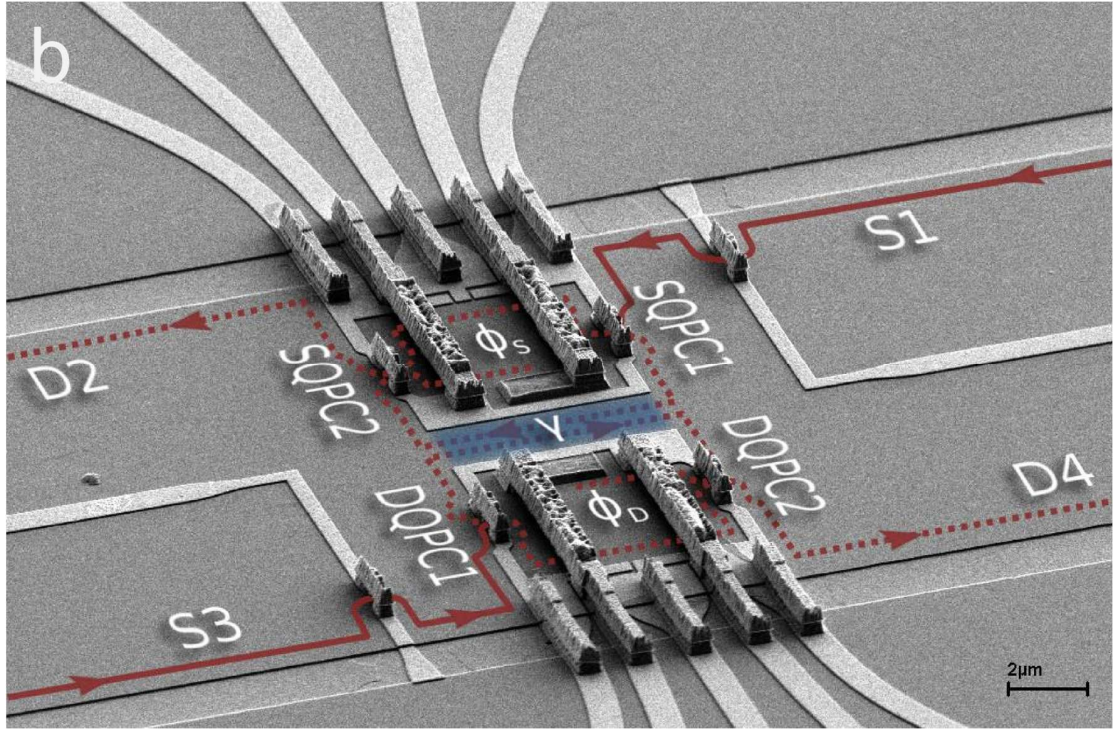

**Figure 1:** Schematics and micrograph of the device under study. (a) The electronic quantum eraser consists of two identical electronic Mach-Zehnder interferometers (MZIs) entangled via Coulomb interactions. Each MZI interferes the outer edge channel in filling factor 2 of the integer quantum Hall effect (lowest spin split Landau level). Quantum point contacts (QPCs) serve as beam splitters, transmitting and reflecting electrons (full lines represent unpartitioned beams; dashed lines represent partitioned beams). Drain currents at each MZI output depend on the Aharonov-Bohm phase due to enclosed magnetic flux and on the interaction with the nearby MZI. (b) A SEM micrograph of the fabricated structure, which was realized in a GaAs-AlGaAs heterostructure harboring a high mobility two-dimensional electron gas. The edge channels were manipulated by biasing surface gates (bright grey) and surface etching. Ohmic contacts serve as sources (S1, S3) and drains (D2, D4), allowing electric access to the electron gas lying underneath the surface. The nano-structures were defined using electron-beam lithography. The device was measured in a dilution refrigerator at a temperature of 12 mK and a magnetic field of 4.4 T.

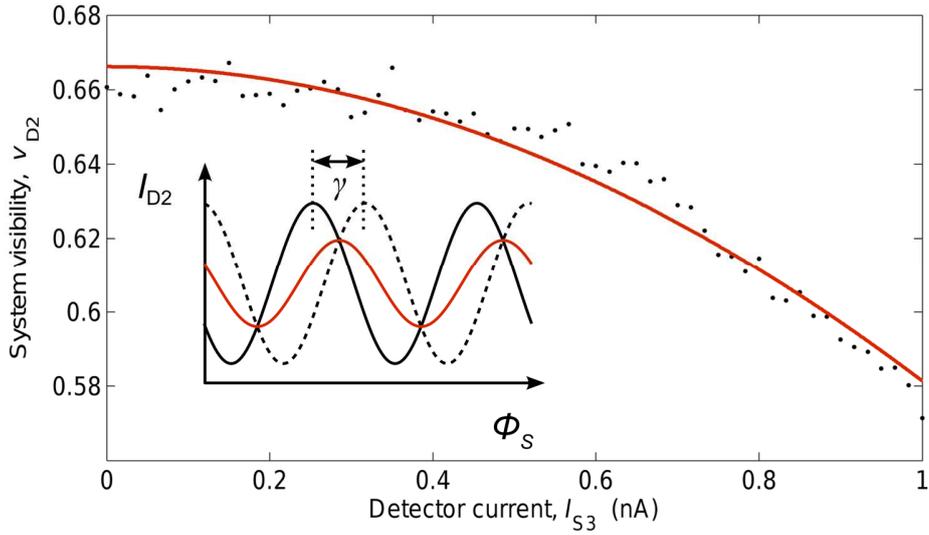

**Figure 2:** Dephasing of the System by entanglement with the Detector. The System visibility is reduced as the Detector current is increased (dots). Coupling the Detector to the System's path acquires *which-path* (WP) information via a phase shift γ in the Detector's phase, which is proportional to Detector current $I_{S3}$. Alternatively, the partitioned current in the Detector's interacting path (carrying shot noise) 'back acts' on the System, thus inducing randomly occurring γ phase shifts. (Inset) The Aharonov-Bohm (AB) dependent current at D2 is shown separately for electrons that did (dashed line) and did not (solid line) receive a "γ shift". The average (red line) shows AB oscillations with the same periodicity but reduced amplitude (partial dephasing). The functional dependence of the System visibility (red solid line) fits with Eq. 6, assuming $\gamma = \pi/6$ at $I_{S3}$=1 nA.

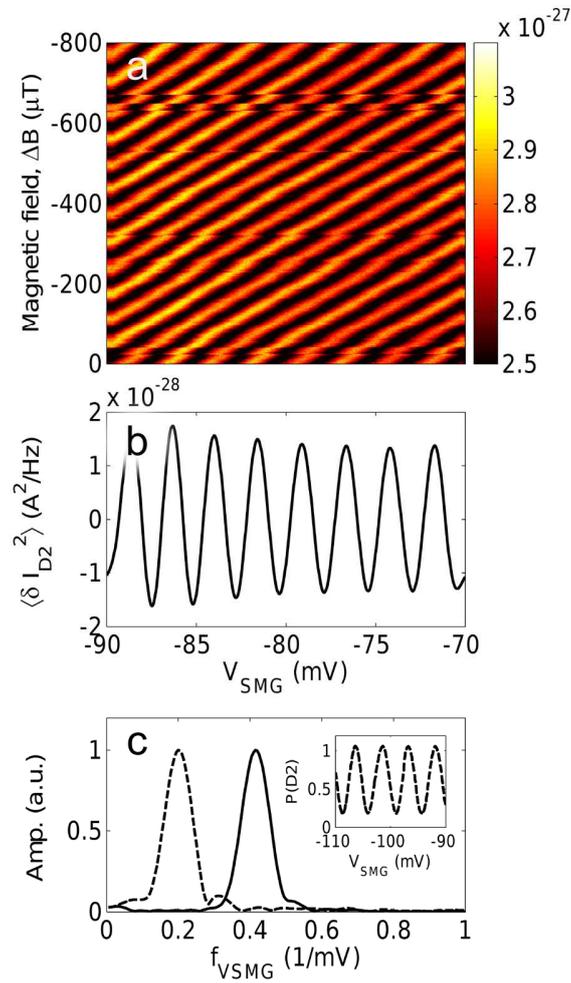

**Figure 3:** AB dependence of auto-correlation (AC) in the System. (a) The System's AC was measured letting the magnetic field decay slowly (Δ*B*), while varying the System's area with the modulation gate (SMG) faster. (b) The average auto-correlation signal oscillates at half the periodicity of the AB oscillation in conductance, as seen in (c) a discrete Fourier transforms of the AB oscillations in conductance (dashed line, 0.2 mV$^{-1}$) and in auto-correlation (full line, 0.4 mV$^{-1}$). Inset: AB oscillations in conductance.

14

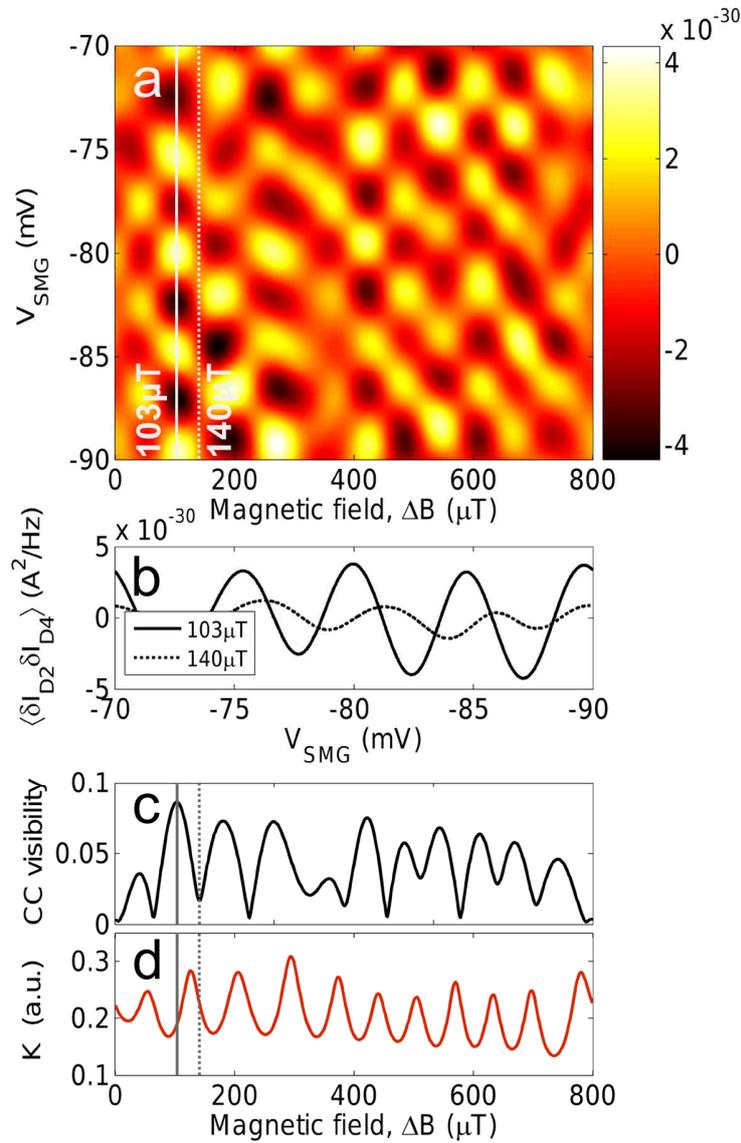

**Figure 4:** Phase recovery by quantum erasure. (a) The cross-correlation (CC) of the System and Detector output currents. (b) Changing the System's flux via SMG produces AB oscillations of the CC at the same frequency as that of the AB oscillation of the conductance. The amplitude of the reconstructed oscillation is modulated by the magnetic flux in the detector: when WP information has been 'erased' the oscillation amplitude is maximal (solid line) and when WP knowledge is maximal the oscillation is minimal (dashed line). The near anti-correlation between (c) the oscillation visibility and (d) the knowledge is a manifestation of the complementarity principle: having access to a particle-like property of the System, via WP information, prevents observing its wave-like nature, such as interference, and vice versa.

15


\maketitle

\section*{Detailed treatment}
  \subsection*{The Electronic Mach-Zehnder Interferometer}
    Manipulating the geometry of the 2DEG, either by permanent etching or by electrostatic depletion, provides control over the shape of the edge states. In particular, bringing two opposite edge states close together over a short distance, in a structure known as a \term{quantum point contact (QPC)}, allows controlled tunneling between the edges.  

    The properties of a QPC, which is the electronic equivalent of an optical beam splitter, are described by a scattering matrix 
    \[ S=  
      \begin{pmatrix}
      r & it \\
      it &r
      \end{pmatrix}
    \]
  where \({t,r}\in\mathbb{R}, \abs{r}^2+\abs{t}^2=1\) are the QPC's transmission and reflection amplitudes, respectively. The scattering matrix links the amplitudes of the two states entering the QPC to those of the two states exiting it. 
   
  An electronic Mach-Zehnder interferometer (MZI) is composed of two QPCs in succession 
  The first QPC puts an impinging electron into a superpositions of being transmitted and being reflected. The two paths are recombined at the second QPC, the resulting interference reflecting the phase difference between the two paths taken. The accumulated phase \ct{\phi} depends on the magnetic flux enclosed by the two paths.
  
  \begin{figure}
    \centering
    \includegraphics[width=0.6\textwidth]{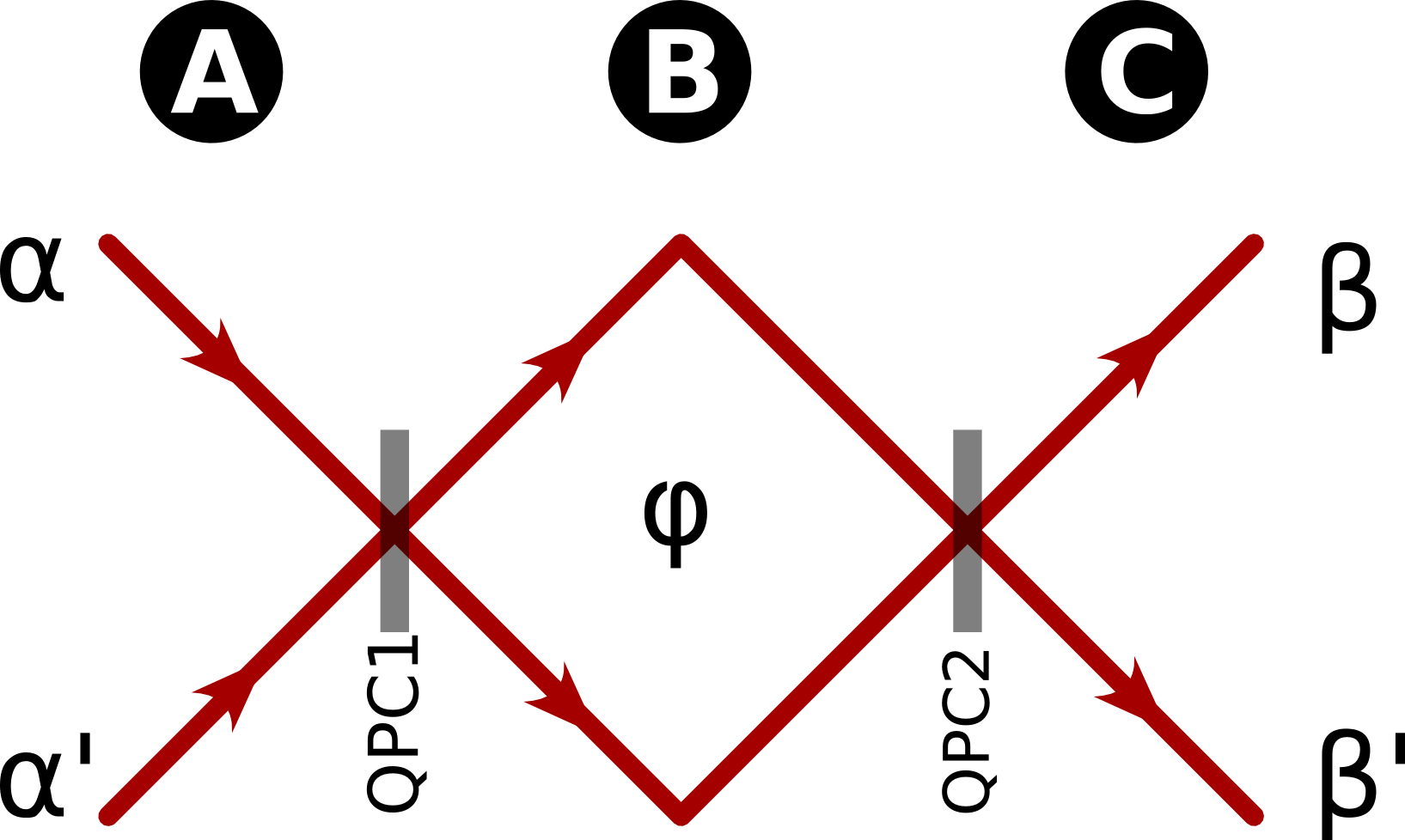} 
    \caption{An electronic Mach-Zehnder interferometer (MZI). The electronic MZI is the electronic counterpart of the optical MZI, where two potential barriers take the role of bean splitters. An electron emanating from a source contact impinges upon the first barrier (Region A) and is put into a superposition of being in the lower and upper arms (Region B). The two arms reunite at the second barrier, allowing the two states to interfere. The current at each drain contact depends on the phase difference between the two paths (Region C).}
    \label{fig:TheoryMZI}
  \end{figure}

  Let us denote the portions before, between and after the two QPCs with A,B, and C, respectively. The Hilbert space before the QPCs (region A) is spanned by the two source states
  \[\ket{\alpha}\equiv\colvec{1}{0},\ket{\alpha'}\equiv\colvec{0}{1};\]
  The intermediate region is spanned by the upper and lower path states, 
  \[\ket{\uparrow}\equiv\colvec{1}{0},\ket{\downarrow}\equiv\colvec{0}{1};\]
  The region after the QPCs is spanned by the detector states
  \[\ket{\beta}\equiv\colvec{1}{0},\ket{\beta'}\equiv\colvec{0}{1}.\]
  In all of our experiments the initial state of the system is \ct{\ket{\alpha}} and the final state is \ct{\ket{\beta}}.
  
  To obtain the system's state between the QPCs (region B) we apply SQPC1's scattering matrix to the initial state:
  \[
   \ket{S_B}=S_1\ket{\alpha}=\colvec{r_1}{-it_1}.
  \]
  In order to get state of the system after interfering at SQPC2 (region C) we apply an operator accounting for the AB phase \ct{\phi_S} accumulated in region B,
  \[
   S_{\phi_S}= 
   \begin{pmatrix}
   e^{i\phi_S} & 0 \\
   0 & 1
   \end{pmatrix},
  \]
  and the scattering matrix for SQPC2:
  \[
   \ket{S_C}=S_2S_{\phi_S}\ket{S_B}=\colvec{r_1r_2e^{i\phi_S}-t_1t_2}{ir_1t_2e^{i\phi_S}+it_1r_2}.
  \]
  The overall probability for an electron to be transmitted into state \ct{\ket{\beta}} is simply
  \[
   P(\beta)=\abs{\braket{\beta}{S_C}}^2=T_0-T_1\cos(\phi_S)
  \]
  where \(T_0=\left(t_1t_2\right)^2+\left(r_1r_2\right)^2, T_1=2t_1t_2r_1r_2\). Since consecutive tunneling events of two electrons are statically independent, the total current at \ct{\beta} depends linearly on the input current: \(I_\beta=P(\beta)I_\alpha\). The visibility of the MZI is defined as 
  \[
   v_\beta = \frac{max(I_\beta)-min(I_\beta)}{max(I_\beta)+min(I_\beta)}=\xi\frac{T_1}{T_0}
  \]
  where \(0\leq\xi\leq 1\) is a phenomenological constant describing the loss of visibility due to unwanted dephasing by the environment.

\subsection{Entanglement and Quantum Erasure}
  \subsubsection*{Entanglement and dephasing}
	  Consider a complex of two MZIs formed using different edge channels and positioned such that the two reflected arms each MZI are in close proximity. Electrons passing simultaneously in the neighboring paths will interact electrically, rendering the two MZI entangled. We make an arbitrary choice to call one of these MZI the system, S, and the other the detector, D. Each is sectioned, as above, into the region before, in-between, and after the QPCs (regions A,B,C). The electrostatic interaction, when it takes place, changes the detector electron's trajectory, adding \term{\ct{\gamma}} to its phase. That is, if the system is \ct{\ket{\downarrow}_S} the state of the detector in region B would be
	  \[
	   \ket{0}_D=S_3\ket{\alpha}_D=\colvec{r_3}{it_3}_D
	  \]
	  and if the system is \ct{\ket{\uparrow}_S} the state of the detector would be
	  \[
	   \ket{\gamma}_D=S_\gamma S_3\ket{\alpha}_D=\colvec{r_3e^{i\gamma}}{it_3}_D.
	  \] 
	  The system-detector complex is in an entangled state, which is a superposition of the above mentioned options:
	  \[
	   \ket{\Psi_B}=r_1\ket{\downarrow}_S\otimes\ket{0}_D+it_1\ket{\uparrow_S}\otimes \ket{\gamma}_D.
	  \]
	  
	  The transmission probability is found in the same manner as for a single MZI. We advance the state \ct{\ket{\Psi_B}} to region C:
	  \begin{align*}
	   \ket{\Psi_C} &= S_4S_{\phi_D}S_2S_{\phi_S}\ket{\Psi_B} \\
					&=\colvec{r_1r_2e^{i\phi_S}}{ir_1t_2e^{i\phi_S}}_S\otimes\colvec{r_3r_4e^{i(\phi_D+\gamma)}-t_3t_4}{ir_3t_4e^{i(\phi_D+\gamma)}+it_3r_4}_D+\\
					&\quad \colvec{-t_1t_2}{it_1r_2}_S\otimes\colvec{r_3r_4e^{i\phi_D}-t_3t_4}{ir_3t_4e^{i\phi_D}+it_3r_4}_D.
	  \end{align*}
	  The transmission of the system MZI is altered by the interaction:
	  \begin{align*}
		P(\beta_S)&=\abs{\braket{\beta_S}{\Psi_C}}^2=T_{0,S}-T_{1,S}Re\left(e^{i\phi_S}\braket{0}{\gamma}_D\right)\\
		&=T_{0,S}-T_{1,S}\left({r_3}^2\cos(\phi_S+\gamma)+{t_3}^2\cos(\phi_S)\right)
	  \end{align*}
	  where \(T_{0,S}=\left(t_1t_2\right)^2+\left(r_1r_2\right)^2, T_{1,S}=2t_1t_2r_1r_2\).
	  Not surprisingly, the transmission of the detector is similar:
	  \[
		P(\beta_D)=\abs{\braket{\beta_D}{\Psi_C}}^2=T_{0,D}-2T_{1,D}\left({r_1}^2\cos(\phi_D+\gamma)+{t_1}^2\cos(\phi_D)\right)
	  \]
	  where \(T_{0,D}=\left(t_3t_4\right)^2+\left(r_3r_4\right)^2, T_{1,S}=2t_3t_4r_3r_4\).
	  
  \subsubsection*{Quantum Erasure and Recovery}	
	  The joint probability for two events, that is the probability for both to happen, is found by taking the expectation value for the product of their projection operators. Specifically, the joint probability of detecting electrons at \ct{\beta_S} and \ct{\beta_D} is
	  \begin{align*}
		P(\beta_S \beta_D)&=\langle\beta_S \beta_D\rangle=\abs{\braket{\Psi}{\beta_S}\braket{\beta_S}{\beta_D}\braket{\beta_D}{\Psi}}\\
		&=\abs{\bra{\beta_D}\beta_S\ket{\Psi}}^2.
	  \end{align*}
	  If the events at the drains are independent the joint-probability is simply \(P(\beta_S \beta_D)=P(\beta_S)P(\beta_D)\). We are interested in the reduced joint probability, \(P(\Delta\beta_S \Delta\beta_D)=P(\beta_S \beta_D)-P(\beta_S)P(\beta_D)\), which is the non-trivial correlation between the two drains, originating in the system-detector entanglement.

	  In the general case, the expression for the reduced joint probability is a rather cumbersome one
	  \begin{align*}
		P(\Delta\beta_S \Delta\beta_D)&=P(\beta_S \beta_D)-P(\beta_S)P(\beta_D)=\\
			&\quad\abs{r_1r_2e^{i\phi_S}\left(r_3r_4e^{i(\phi_S+\gamma)}-t_3t_4\right)-t_1t_2\left(r_3r_4e^{i\phi_S}-t_3t_4\right)}^2-\\
			&\quad\left[(t_1t_2)^2+(r_1r_2)^2-2r_1r_2t_1t_2\left({r_3}^2\cos(\phi_S+\gamma)+{t_3}^2\cos(\phi_S)\right)\right]\times\\
			&\quad\left[(t_3t_4)^2+(r_3r_4)^2-2r_3r_4t_3t_4\left({r_1}^2\cos(\phi_D+\gamma)+{t_1}^2\cos(\phi_D)\right)\right].
	  \end{align*}
	  But for symmetrically tuned MZIs (\(t_i,r_i=\tfrac{1}{\sqrt{2}}, i=1\ldots 4\)), the above reduces to 
	  \[
	  P(\Delta\beta_S \Delta\beta_D)=\tfrac{1}{4}\cos\left(\phi_S+\frac{\gamma}{2}\right)\cos\left(\phi_D+\frac{\gamma}{2}\right)\sin^2\left(\frac{\gamma}{2}\right).
	  \]
	  
	  The autocorrelations \ct{(P(\Delta\beta^2)} can be obtained in a similar manner, taking note that \(P(\beta\beta)=P(\beta)\):
	  \[
	  P(\Delta\beta_j\Delta\beta_j)=P(\beta_j)\left(1-P(\beta_j)\right), j=S,D
	  \]
	  which determines the power spectral density at the MZI drain:
	  \begin{align*}
		S_\beta&=2eI_\alpha P(\beta) \left(1-P(\beta)\right)\\
		&= T_0(1-T_0)+T_1(2T_0-1)\xi\cos(\phi)-{T_1}^2\xi^2\cos^2(\phi).
	  \end{align*}

\section*{Technical considerations}
	\subsection{Biasing limitation}
	The currents used in our setup (and similar setups) were limited to about 1 nA per edge channel. Higher currents were observed to tunneling to adjacent co-propagating edge channels. Note that even for smaller currents and counter-propagating edge channel we observed a minute signal transfer (\textless 1\%) between the edges  at high-frequency, probably due to capacitive coupling. In addition, most QPCs show bias dependent transmission profile, especially at higher voltage biases. In our experiment we took care to work in a regime where no significant current tunneling occurred and QPC transmission was not bias dependent. 
	\subsection{Random phase noise} 
	During our experiment we observed that the interference of a MZI constructed using the innermost edge channel is plagued by random phase noise: both QPCs forming the MZI showed stable transmission, the interference had high visibility, comparable to other edges channels, but the AB phase jumped randomly at a rate of about 1 Hz. Our understanding is that these jumps originate from random charging and discharging events (e.g. in quantum dots formed in the bulk due to crystal defects) in the vicinity of the MZI, which is highly sensitive to charge fluctuations, especially to those occurring in nearly isolated quantum dots \cite{weisz_2012}. However, when interfering the outer edge channels, the innermost edge channel engulfs the interfering edge, shielding it from the above-mentioned charge fluctuations, thus eliminating the random phase noise. For this reason, it was essential in our experiment to let the inner edge pass through the interaction region, even at the expense of weakening the interaction, both due to increasing the distance between the interacting edges and the shielding by the inner edge channel.
	\subsection{MZI lobe structure}
	The MZI transmission is energy dependent in a non-trivial way \cite{neder_2006}, declining as the source-drain bias is increased; when it reaches zero, it recovers, with a \(\pi\) phase shift and then declines again, forming a so-called 'lobe pattern'. For a DC, the overall AB oscillation amplitude is set by a summation on all energies, up to the applied voltage bias. Therefore, the maximal visibility is obtained at the voltage bias for which the energy-dependent transmission reaches zero. For our MZI this occurred at about 6.5 \ct{\mu eV}, which implies a current of 0.5 nA. 

\section*{Other configurations examined}
	We attempted several different configurations in hopes of increasing the maximal interaction between the two MZI. 
	\subsection{Gate separated double MZI}
	One approach form two MZI where the interacting paths are separated by a very thin (50 nm) metallic gate. We hoped that the reduced distance, and the absence of shielding by the inner edges will boost the interaction. Although the MZI showed very high visibility (\textgreater 90\%), the interaction was too small to observe. We attribute this to the separating gate's ability to screen the two MZI from each other. 
	\begin{figure}
		\centering
		\includegraphics[width=0.6\textwidth]{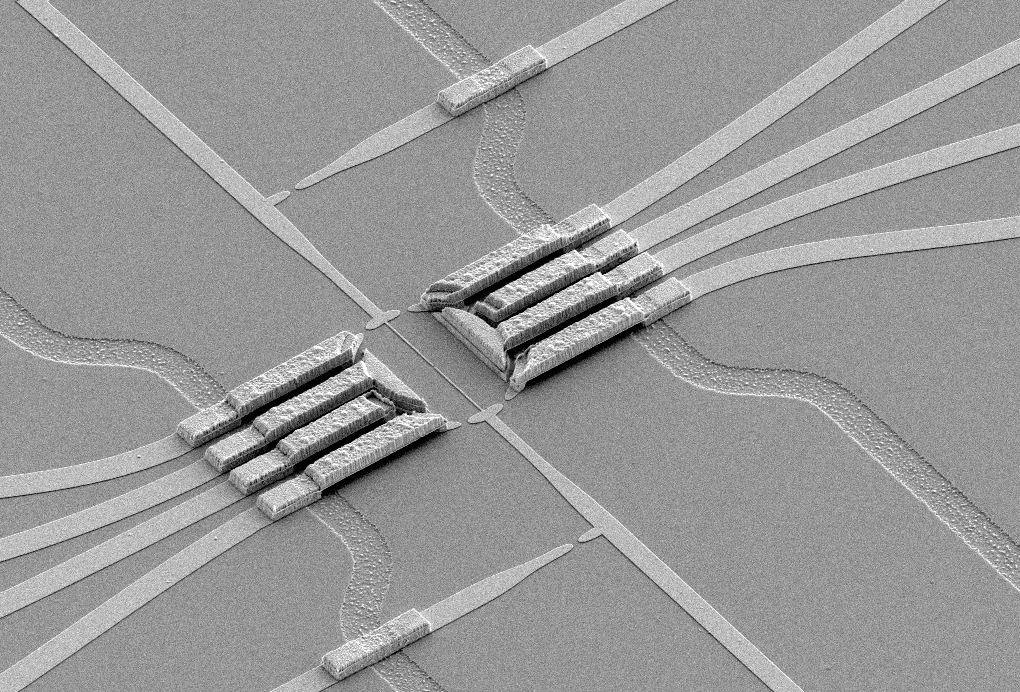} 
		\caption{An alternative EQE setup. The interacting paths are separated by a thin surface gate, in hopes of achieving stronger interaction. However, the gate seems to be very efficient in shielding the two MZI from each other.}
		\label{fig:TheoryMZI}
	\end{figure}   
	\subsection{Co-propagating edges double MZI}
	Another approach was to construct the two MZI using different, co-propagating edges. Since the co-propagating edges are very close and there is nothing to screen them from each other, we expected to observe a strong interaction. For this purpose we filling factor 3 of the IQHE: the outermost edge formed one MZI, the middle edge formed another MZI and the innermost edge shielded the two MZI random charge fluctuations in the bulk (see above). Indeed, this setup achieved strong interaction between the two MZI, gaining full dephasing for currents of 1.0 nA per edge channel. However, it turned out that the tunneling between co-propagating edges was large to the extent of preventing useful measurements. The tunneling occurred only when the MZI were tuned, hinting to some resonative source for this behavior.

\bibliographystyle{unsrt}
\bibliography{Biblio_General}

\begin{thebibliography}{1}

\bibitem{weisz_2012}
E~Weisz, HK~Choi, M~Heiblum, Yuval Gefen, V~Umansky, and D~Mahalu.
\newblock Controlled dephasing of an electron interferometer with a path
  detector at equilibrium.
\newblock {\em Phys. Rev. Lett.}, 109(25):250401, 2012.

\bibitem{neder_2006}
I.~Neder, M.~Heiblum, Y.~Levinson, D.~Mahalu, and V.~Umansky.
\newblock Unexpected behavior in a two path electron interferometer.
\newblock {\em Phys. Rev. Lett.}, 96, 2006.

\end{thebibliography}

\end{document}